# Choice of tip, signal stability and practical aspects of Piezoresponse-Force-Microscopy

L. F. Henrichs, J. Bennett, A. J. Bell

*Institute for Materials Research, University of Leeds, Engineering Building , LS2 9JT Leeds, United Kingdom*

Piezoresponse force-microscopy (PFM) has become the standard tool to investigate ferroelectrics on the micro- and nanoscale. However, reliability of PFM signals is often problematic and their quantification is challenging and thus not widely applied.

Here, we present a study of the reproducibility of PFM signals and of the so-called PFM background signal which has been reported in literature. We find that PFM signals are generally reproducible to certain extents. The PFM signal difference between 180° domains on periodically-poled lithium niobate (PPLN) is taken as the reference signal in a large number of measurements, carried out in a low frequency regime (30-70 kHz). We show that in comparison to Pt coated tips, diamond coated tips exhibit improved signal stability, lower background signal and less imaging artifacts related to PFM which is reflected in the spread of measurements. This is attributed to the improved mechanical stability of the conductive layer. The average deviation of the mean PFM signal is 38.3%, for a diamond coated tip. Although this deviation is relatively high, it is far better than values from literature which showed a deviation of approx. 73.1%. Additionally, we find that the average deviation of the background signal from 0 is 11.6% of the PPLN domain contrast. Thus, the background signal needs to be taken into account when quantifying PFM signals and should be subtracted from PFM signals. Those results are important for quantification of PFM signals, since PPLN might be used for this purpose when PFM signals measured on PPLN are related to its macroscopic $d_{33}$ coefficient. Finally, the crucial influence of sample polishing on PFM signals is shown and we recommend to use a multistep polishing route with a final step involving 200 nm sized colloidal silica particles.

## I. INTRODUCTION

Over the past years, the number of publications with relation to PFM has grown constantly and it has become the standard technique to study ferroelectrics on the nanoscale.[1-7] However, it has often been a matter of debate to what extent the magnitude of PFM signals can be reproduced and quantified experimentally. Some researchers express doubts whether the magnitude of PFM signals of two different experiments can be reasonably reproduced and refer to PFM measurements taken on PPLN on different days which are shown in Fig. 1.[8]

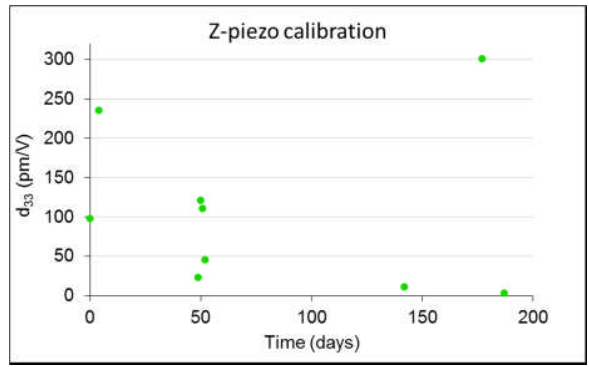

FIG. 1 Piezoelectric coefficients for PPLN measured by PFM on different dates using the z-piezo method. Graph built from data taken from literature in the low frequency range.[8]

The range of measured values on PPLN on different days is from 3 to 301 pm/V while the average deviation of the mean is 76.9%. These measurements would in fact suggest that reproducibility of PFM measurements is hardly possible. However, these PFM signals had been calibrated using a technique which we will call z-piezo method henceforth and we propose that the large deviation is at least partly due to the inaccuracy of this method as we will argue below.

The z-piezo method is currently one of the most widely used techniques of quantification in PFM since it is standardly implemented in AFMs by Asylum Research®, which are widely used for PFM. It involves the z-piezo element of the AFM.[9, 10] The principle is depicted schematically in Fig. 2.

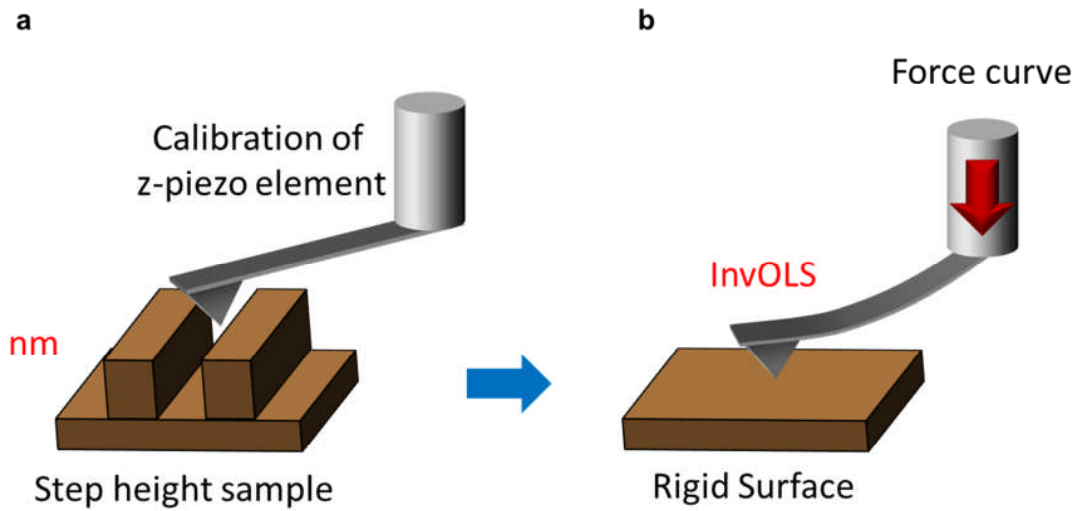

FIG. 2 Schematic illustration of the z-piezo method

Initially, this method requires calibration of the z-piezo element e.g. by using a height-calibration sample (Fig. 2a). Afterwards, the so called inverse optical lever sensitivity (InvOLS) which connects cantilever deflection to a certain height difference is obtained. This can be extracted from the slope of the repulsive part of a force curve taken on the sample. With this method, cantilever deflections can in principle be related to sample deformations due to the converse piezoelectric effect and thus PFM signals can be quantified.

However, a major drawback of this method is, that force curves are taken at much lower frequencies as compared to cantilever deflections in PFM experiments which are in the range of several 10 kHz up to several MHz. This discrepancy inevitably results in a calibration error, e.g. due to the frequency dependence of electronic components of the AFM system such as the four-quadrant photodiode or electric amplifiers. The fact that the initial height information ultimately used for PFM-signal calibration is obtained on height-calibration samples which are usually 2-3 orders of magnitude larger than actual sample deformations measured in PFM, adds another factor of uncertainty. Furthermore, height-calibration involves the use of the AFM's feedback loop, which is not the case in a PFM experiment.

A more accurate method for quantifying PFM signals might be to use a ferroelectric reference sample. Such an approach has the advantage that conditions during calibration (frequency, magnitude of deformation) are very similar as during the actual measurement, which would eliminate several errors with respect to the z-piezo method. However, one needs to take into account an inherent background signal in PFM which has been reported by Soergel et al[1] which can vary with frequency and presumably depends on the condition of the tip but is independent of the sample. This background signal can even be measured on non-piezoelectric materials such as glass or metal. Due to this fact, a single crystalline reference sample, containing only ferroelectric domains with equal magnitude but different sign of polarization, (i.e. only 180° domains), can be used to calibrate PFM signals and measure the background signal at the same time. Because the background signal is equal for both up and down domain, the level of the background signal is simply the mean of the two signals. One material that fulfils those requirements is periodically-poled lithium niobate (PPLN). It is readily available and robust to e.g. changes in temperature. Crystals are usually cut perpendicular to the Z-direction, so that measurements can be carried out on the Z-face. Hence, PFM signals can be calibrated when the signal difference for up and down domains is related to twice the macroscopic value of the $d_{33}$ coefficient of PPLN which is approx. $d_{33} = 20$ pm/V.[11, 12]

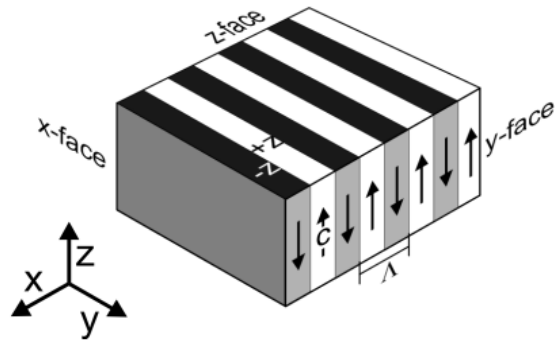

FIG. 3 Schematic of PPLN crystal showing crystal axes and polarization directions.

Here, we investigate the statistical deviation of PFM signals measured on PPLN on different days similarly as shown in Fig. 1 to assess the reproducibility of PFM signals. Furthermore, we also measure the background signal at the same time to assess its magnitude and influence on PFM signals. Our results are important to answer the question, to what extent PFM signals are reproducible at all. But also for the prospect of using PPLN as a reference sample, reproducibility is important to get an idea on the stability of calibration over a certain time.

Furthermore, we will also address common practical issues like imaging artifacts and sample polishing which affect reliable and reproducible PFM data acquisition, and give advice on how to avoid these problems.

## II. EXPERIMENTAL

PFM measurements were carried out using a 5420 Atomic-Force Microscope by Agilent® on one sample of a commercial PPLN single crystal. Two different types of AFM tips were used: one type was coated with a conductive layer of Pt on top of a Ti layer while the other tips were coated with a conductive layer of nitrogen doped diamond. For each tip types, always the same make and model was used. Both types of cantilevers were made of doped silicon with typical force constants of approx. 5 N/m.

The background signal has consequences for the measurement of PFM signals and makes acquisition of the X-channel signal of the lock-in-amplifier (LIA) more beneficial than acquisition of Amplitude and Phase.[13] Therefore, the difference in the X-channel between +Z and –Z domain (up and down domain) of PPLN, was recorded which was usually averaged over 256 lines. A +Z and –Z domain contrast in the Y-channel was usually minimized by adding a phase-shift to the LIA signal with respect to the drive signal. However, the contrast in the Y-channel was always small compared to contrast in X-channel as expected in PFM due to the fact that signals should in principle be always in-phase or 180° out of phase with respect to the driving frequency.[13]

All measurements were done with an AC drive voltage of ±10 V and with the same amplification value of the LIA (if different amplification was used, values were converted accordingly). It should be noted that the use of lower drive voltages around ±0.5-1 V is generally favorable, since the use of high drive voltages can result in problems such as a phase lag resulting in a background signal. This is especially important when measuring on thin-films. However, when using thick ferroelectric samples such as bulk ceramics or single crystals, the electric fields involved are generally smaller as compared to very thin samples[1]. The drive frequency was in the range of 30-70 kHz. However, in this "low-frequency" range the instrument did not have a large frequency dependence of the PFM signal.

No specific condition of the tip was maintained, i.e.values were not necessarily recorded with a new or sharp tip. The tips used for the measurements might have been new or used but always showed good or reasonable resolution and signal-to-noise ratio. Usually values would stay relatively constant during the same day when multiple measurements were carried out on that day. Only one value was picked for a single day to assess the variation of PFM signals over longer timescales.

PFM images presented in Fig. 6 and Fig. 7 were recorded with similar settings as above. The sample shown in Fig. 7 c, d was polished in several steps using various polishing-cloths (TexMet P®, TriDent®, ChemoMet® from Buehler) in combination with diamond abrasive-liquids where the diamond particle-size was gradually reduced for consecutive steps (9 μm, 3 μm, 1 μm), until a final polishing step involving 200 nm sized colloidal silica particles was reached.

## III. RESULTS AND DISCUSSION

### A. PFM signal stability

Fig. 4 illustrates how values for the assessment of PFM signal stability and background signal were obtained.

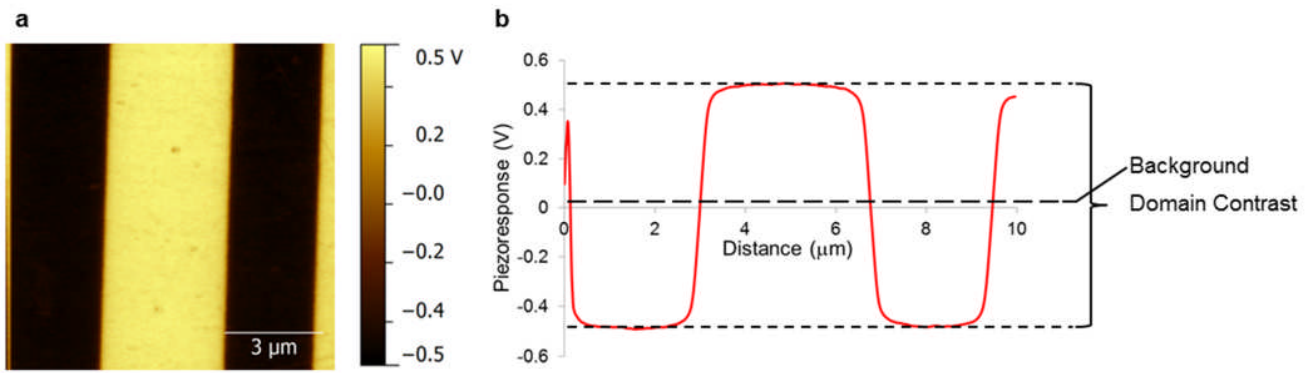

FIG. 4. (a) A typical vertical PFM image (X-channel or mixed signal) of PPLN. (b) Cross-section of image with domain contrast and background signal.

A typical PFM image (X-channel or mixed signal) is shown in Fig. 4a. Fig. 4b shows the averaged cross-section of this image with domain contrast and background signal. These values were collected for many measurements and are displayed in graphs as illustrated in Fig. 5.

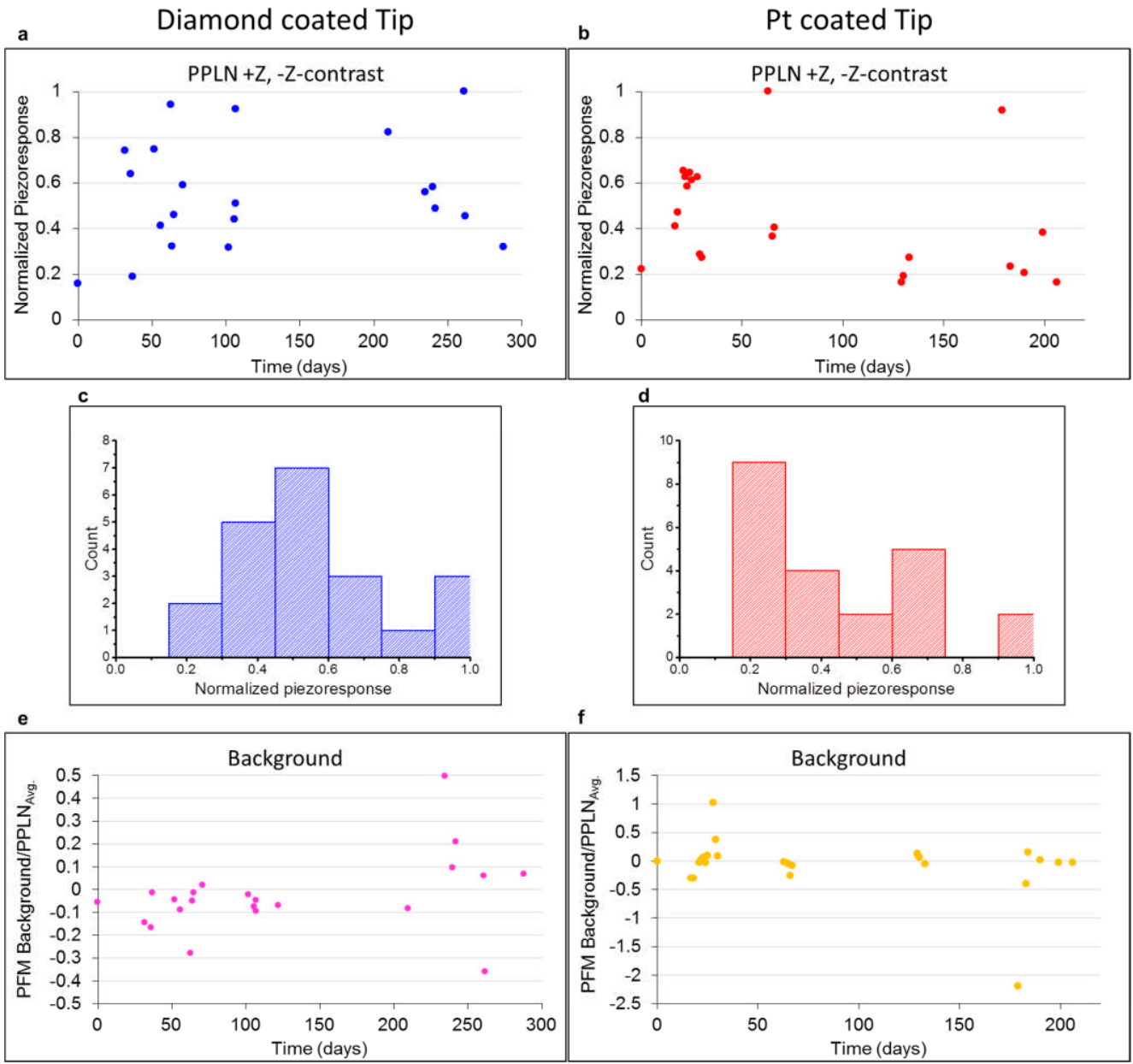

FIG. 5 Vertical PFM contrast of +Z to –Z domains of PPLN measured with a conductive diamond coated tip (a) and a Pt coated tip (b). (c) and (d) show histograms corresponding to (a) and (b) respectively. PFM background signal divided by average +Z to –Z PPLN domain contrast (PPLN$_{Avg.}$) for conductive diamond coated tip (e) and a Pt coated tip (f).

Overall, signals varied between 0.26 V and 2.40 V (Output voltage of LIA) with average deviations of 38.3% and 48.8% of the mean for diamond and Pt coated tips respectively. Both tips had similar values of the average output voltage of the LIA of close to 1 V which indicates that the average piezoresponse was similar for both tips.

It is important to note that the dispersity of values for diamond and Pt coated tips differ signicantly (Fig. 5 a and b). In case of diamond coated tips, the dispersity of values is rather uniform, whereas for Pt coated tips, there are few very high values and the majority of values being in the range between 10-50% of the highest (see histograms in Fig. 5 c and d respectively). We propose that the large values correspond to an intact conductive layer while the lower ones correspond to worn tips with degraded conducting layer. Since both new and used tips were used randomly for both tip types, it seems that Pt coated tips degrade much faster which results in the majority of measurements having low values, while diamond coated tips seem to always remain very stable. This is in line with the observation, that the resolution of new Pt coated PFM tips decreased very rapidly (approx. a few scans with 5 x 5 μm scan size and 256 lines) during imaging, whereas this is not the case for diamond coated tips. Of course, this is expected from the mechanical properties of the two film materials. Fast degradation of the Pt layer is probably due to scanning in contact mode and the relatively stiff cantilevers used for PFM. This diminishes the advantage of lower tip diameter of Pt coated tips (approx. 25 nm) in comparison to diamond coated tips (approx. 100 nm).

Fig. 5 f shows that the background signal can vary strongly from -218.9% up to 102.5% of the average PPLN domain contrast for the platinum coated tip, whereas for the diamond coating the range lies only between -35.8% and 49.9% (see Fig. 5 e). While the average background signal for many measurements is relatively close to 0 for both tips, the average deviations from 0 are 11.6% and 24.0% of the mean PPLN domain contrast for diamond and Pt coated tips respectively. This shows that the background signal usually cannot be neglected for samples with a similar piezoresponse as PPLN which is the case for many ferroelectric samples. If the background signal is not corrected, it distorts PFM signals (especially Amplitude and Phase) as described in literature and therefore, should be corrected by determining the background signal and subtracting it from X-channel data.[13] The above results are summarized in Table 1.

TABLE I Comparison of results for different conditions.

| Technique | PPLN | | z-Piezo[8] |
|---|---|---|---|
| **Tip** | Cond. Diamond | Pt on Ti | - |
| **Min, Max Value for PPLN contrast** | 0.26, 1.6 V | 0.34, 2.4 V | 3, 301 pm/V |
| **Mean Value** | 0.97 V | 1.03 V | 105 pm/V |
| **Average Dev. of Mean** | 38.3% | 48.8% | 73.1% |
| **Min, Max Value of Background/PPLN$_{Avg.}$** | -35.8%, 49.9% | -218.9%, 102.5% | - |
| **Mean Value of Background/PPLN$_{Avg.}$** | -2.9% | -7.0% | - |
| **Average Dev. Background/PPLN$_{Avg.}$** | 10.5% | 26.0% | - |

Our results indicate that PFM data are reproducible to a certain extent although signal variation is relatively high. Tips coated with conductive diamond exhibit better signal stability and lower background signals as compared to Pt coated tips, which is reflected in the dispersity of the data. We attribute this to the better stability of diamond coated tips. Although for the above results, always the same manufacturer and model was used for each tip type, comparison to other models indicate that those results are generally applicable for Pt and diamond coated tips.

Furthermore, our results indicate, that the poor reproducibility found by Gruverman et al.[8] is at least partly due to the z-piezo calibration method, which adds another factor of uncertainty to the calibration and thus should not be employed.

### B. Practical Issues

#### 1. Choice of tip

A common artifact in PFM imaging is displayed in Fig. 6.

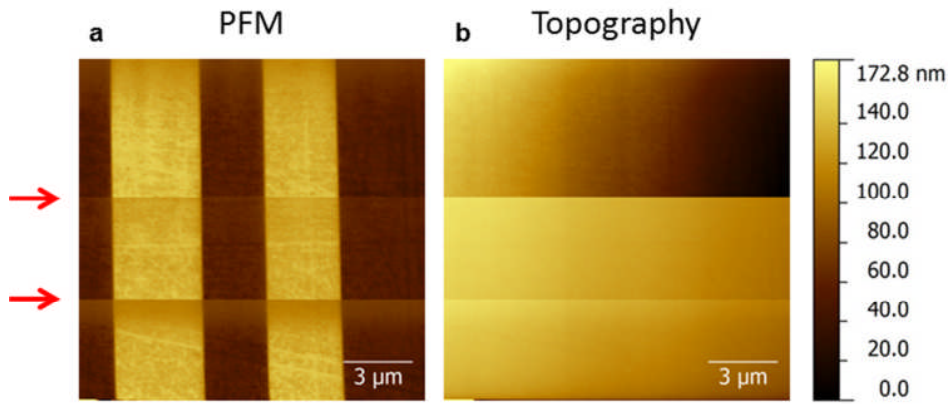

FIG. 6 Common artifact in PFM. "Skips" occur in (vertical) PFM (X-channel or mixed signal) image (a) and topography (b) at the same time (indicated by red arrows).

Here (vertical) PFM signals (X-channel or mixed signal) decrease gradually during scanning (direction of scanning is up) until a "skip" in the signal occurs (Fig. 6a, marked by red arrows). These sudden changes are often accompanied also by a skip in the absolute values of topography which is visible in the unflattened topography image (Fig. 6b). We found that these skips are associated to the PFM experiment because they occurred almost only when an AC voltage was applied. We explain the skips by a sudden release of contaminants on the tip-apex or a sudden change of the conductive layer of the tip which both result in a rapid change of PFM signals due a to change of the electric field distribution at the tip-apex. Furthermore, we found that these skips occurred much more frequently when Pt coated, but not when diamond coated tips were used. This observation is in agreement to the conclusions drawn from the dispersity observed in Fig. 5 a and b. Apparently, the enhanced mechanical stability of the diamond coated tip, also results in less imaging artifacts.  Therefore, we strongly suggest the use of diamond coated tips for PFM.

### *2. Influence of sample polishing*

The influence of appropriate sample polishing on PFM signals is shown in Fig. 7.

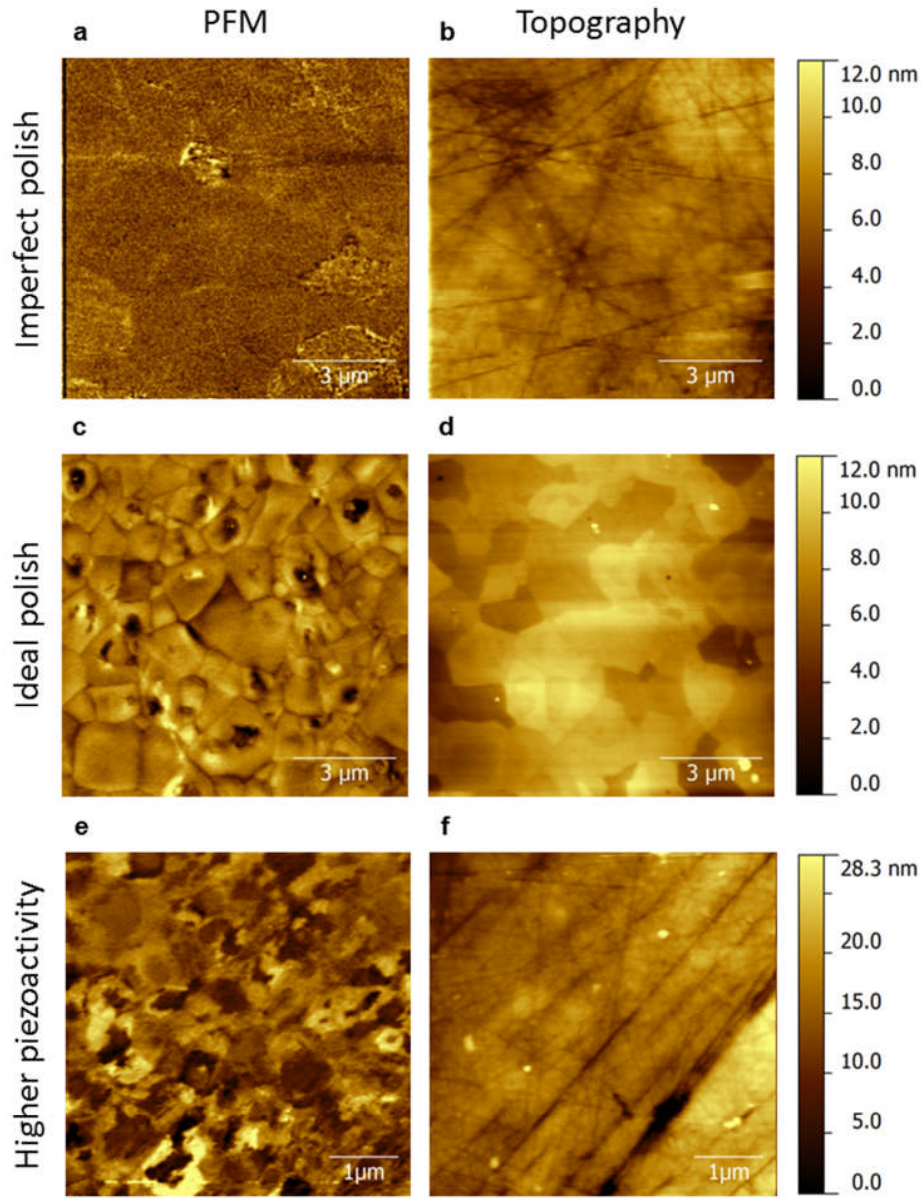

FIG. 7 Influence of polishing on PFM signals. Images (a)-(d) were all obtained on the same material $(BiFe_{0.9}Co_{0.1}O_3)_{0.4}$-$(K_{1/2}Bi_{1/2}TiO_3)_{0.6}$ which has a relatively low $d_{33}$ coefficient. Images (a) and (b) were obtained on a sample that was not sufficiently polished, while images (c) and (d) were obtained on a well-polished sample. PFM images (e) and (f) were taken on a $(BiFeO_3)_{0.65}$-$(PbTiO_3)_{0.35}$ ceramic with relatively high $d_{33}$ coefficient. Here PFM signals are strong even though the sample is not perfectly polished.

Fig. 7a,b and Fig. 7c,d show (vertical) PFM images and topography for an insufficiently and a for a well-polished sample of the same composition, respectively. Note that while in Fig. 7b scratches are visible in topography, this is not the case for Fig. 7d where no scratches are visible and the grain structure is revealed through appropriate polishing. The corresponding PFM images differ largely. While the PFM image corresponding to the insufficiently polished sample (Fig. 7a) is very noisy and barely exhibits any PFM signals, the image corresponding to the well-polished sample (Fig. 7c) exhibits strong signals and a clear ferroelectric pattern. The last polishing step involving 200 nm sized colloidal silica particles is particularly important, since it is responsible for the difference between samples shown in Fig. 7d and Fig. 7b,f.

Images in Fig. 7a and c were all recorded on ceramic samples of the material $(BiFe_{0.9}Co_{0.1}O_3)_{0.4}$-$(K_{1/2}Bi_{1/2}TiO_3)_{0.6}$ which has a low $d_{33}$ coefficient of approx. 16 pm/V. The low $d_{33}$ coefficient and the fact that the aforementioned material is a relaxor ferroelectric, might explain the strong dependence on sample polishing. In contrast, other materials such as $(BiFeO_3)_{0.65}$-$(PbTiO_3)_{0.35}$ ceramics, which exhibit higher piezoelectric coefficients and are classical ferroelectrics, did not show such a strong dependency (see Fig. 7e,f). However, it is always beneficial to measure on well-polished ceramic samples in order to record information of the true microstructure at the same time with PFM images.

## IV. Conclusions

In contrast to data from literature,[8] we could show that PFM signals are generally stable and reproducible to certain extents. We compared PFM signals on one particular sample independently on different days with different kinds of tips. The lowest average deviation of the mean value for the +Z and -Z contrast of periodically poled lithium niobate ($LiNbO_3$, PPLN) was 38.3% when using a tip coated with conductive diamond. In general, we strongly recommend to use tips with conductive diamond coating for PFM, since they exhibit better reproducibility, lower background signal and less imaging artifacts. Pt coated tips in contrast, apparently degrade fast during scanning which eliminates the advantage of higher resolution as compared to diamond coated tips. Furthermore, we found that the average deviation of the background signal per measurement is 11.6% of the average PPLN domain contrast for a diamond coated tip and thus, the background signal cannot be neglected and should be eliminated.

This study is also intended as a practical guidance to users. Our results give an idea, which level of reproducibility users might expect of their PFM data, without having to take special precautions such as always using very new tips. In many cases, interesting results are obtained after hours or days of using the same tip, which means that the tip is not in a new condition anymore at the time of measurement.

With our method, the user might run a calibration measurement after a measurement that delivered good results, using the same AC drive voltage and frequency. For this scenario, our results give the researcher a good idea to which extent measurements are comparable to other measurements which were calibrated in the same way and taken on a different day with different tip.

Finally, we show that appropriate polishing of samples has a crucial influence on PFM signals especially for samples with low piezoelectric coefficients. Therefore, we suggest the use of a multistep polishing route involving 200 nm sized colloidal silica particles as the last step of polishing.

## V.  Acknoledgements

Work was financially supported by the EU Commission under the Marie Curie Initial Training Network "NANOMOTION" (PITN-GA-2011-290158) which is gratefully acknowledged.

L.F.H. thanks E. Soergel from University of Bonn, Germany for many fruitful discussions and M. Fenner from Keysight Technologies (formerly Agilent Technologies) for excellent technical support.

Processing of AFM-images was done using the free AFM software 'Gwyddion'[14] for which excellent user support by D. Nečas is gratefully acknowledged.

10. https://www.asylumresearch.com/Applications/PFMAppNote/PFMAppNote.shtml, (2015).
11. T. Yamada, N. Niizeki and H. Toyoda, Japanese Journal of Applied Physics **6** (2), 151 (1967).
12. G. C. B. G. Bhagavannarayana, K. K. Maurya, B. Kumar, J. Mater. Res. **15** (2005).
13. T. Jungk, Á. Hoffmann and E. Soergel, Journal of Microscopy **227** (1), 72-78 (2007).
14. D. Nečas and P. Klapetek, Cent. Eur. J. Phys. **10** (1), 181-188 (2012).
14